\documentclass[twocolumn,showpacs,preprintnumbers,amsmath,amssymb,floatfix]{revtex4}
\usepackage{graphicx}% Include figure files
\usepackage{dcolumn}% Align table columns on decimal point
\usepackage{bm}% bold math

\newcommand{\beq}{\begin{equation}}
\newcommand{\eeq}{\end{equation}}
\newcommand{\bey}{\begin{eqnarray}}
\newcommand{\eey}{\end{eqnarray}}

\begin{document}

\title{ Anisotropic Quintessence stars}

\author{ Mehedi Kalam}
\email{kalam@iucaa.ernet.in} \affiliation{Department of
Physics, Aliah University, Sector - V , Salt Lake,  Kolkata -
700091, India}

\author{Farook Rahaman}
\email{rahaman@iucaa.ernet.in} \affiliation{Department of
Mathematics, Jadavpur University, Kolkata 700 032, West Bengal,
India}
\author{ Sajahan Molla}
\email{sajahan.phy@gmail.com} \affiliation{Department of Physics,
Aliah University, Sector - V , Salt Lake,  Kolkata, India}
\author{Sk. Monowar Hossein}
\email{sami_milu@yahoo.co.uk} \affiliation{Department of
Mathematics, Aliah University, Sector - V , Salt Lake,  Kolkata -
700091, India}

\date{\today}

\begin{abstract}
We propose a relativistic model for ,cs with the combination of an
anisotropic pressure corresponding to normal matter and a
quintessence dark energy having a characteristic parameter
$\omega_q$ such that $-1<\omega_q< -\frac{1}{3}$. We discuss various
physical features of the model and show that the model satisfies all
the regularity conditions and can provide stable equilibrium
configurations.
\end{abstract}


\maketitle

\section{Introduction}
Compact objects are of great interest for a long time. Theoretical
analysis of superdense stars have been done by several
 authors
\citep{Rahaman2012a,Kalam2012a,Hossein2012,Rahaman2012b,Kalam2012b,Kalam2013,Lobo2006,Bronnikov2006,Egeland2007,Dymnikova2002}.
Ruderman\citep{Ruderman1972} show that nuclear matter may have
anisotropic behaviors at a very high density($ \sim 10^{18}$
Kg/$m^3$).  Consideration of an anisotropic behavior of the compact
star leads to a realistic situation\citep{Varela2010,rahaman2010}.
Anisotropy
 in matter implies radial pressure($p_r$) is not equal to the  tangential pressure ($p_t$). As the density of a strange star
 exceeds the nuclear density, it is obvious that the pressure at the interior should be anisotropic \citep{Bowers1974,Sokolov1980,
 Herrera1992}. It may occur for various reasons like existence of solid core, phase transition, presence of electromagnetic field etc.
In recent past, L.Herrera and N.O. Santos \cite{HS} provided   an
exhaustive
    review on the subject of anisotropic fluids. More recently a comprehensive work on the influence of
     local anisotropy on the structure and evolution of compact object has been studied by  L. Herrera et. al \cite{H}.\\

In recent WMAP measurement conclude that $73\%$ of the universe is $\it{dark ~ energy}$ \citep{Perlmutter1998,Riess2004}.
 Dark Energy theory is the most accepted one to explain the acceleration of the universe. It has some peculiar properties
 such as negative pressure and violation of the energy conditions. Current experimental data shows that pressure to density
 ratio $\omega$ is in the range $-1.38 < \omega < -0.82$. In the present work, we consider models of compact stars containing
  not only ordinary matter but also a quintessence matter having a characteristic parameter $\omega_q$ such that $-1 < \omega_q
  < -\frac{1}{3}$. The presence of dark energy motivates us to consider the existing strange stars are a mixture of both ordinary
   matter and quintessence matter in different proportions. The study of such kind of mixed matter is now an interesting problem
    and some works has already been done on this direction\citep{Lobo2006,Bronnikov2006,Chan2009,Rahaman2012a}.  \\

Some important works was carried out with Krori and Barua
\citep{Krori1975} (KB)metric by Rahaman et al. \citep{Rahaman2012a}
for singularity-free dark energy stars which represents an
anisotropic compact stellar configuration.The same KB spacetime was
 used to study the strange stars by using the cosmological constant as a dark energy source  by Kalam et al. \citep{Kalam2012a}
  and Hossein et al.\citep{Hossein2012}. Moreover by using the same metric Kalam et al.\citep{Kalam2013} presents a interesting
  anisotropic strange star model.The beauty of the model is that the interior physical properties of the star solely depend on
   the central density of the matter distribution.\\

We would like to mention here the very recent work of Kalam et al.
\citep{Kalam2012b} where the authors have  proposed a model for
strange quark stars within the framework of MIT Bag model in Finch
and Skea metric\citep{Finch and Skea1989}. F.S.
Lobo\citep{Lobo2006}, R. Chan et al.\citep{Chan2009},
Egeland\citep{Egeland2007}, Dymnikova\citep{Dymnikova2002} and many
more
 have also studied stellar structure  in  different ways.\\

As mentioned above basically we have considered here a two fluids
model for compact star with quintessence dark energy as one of the
ingredients. Here we consider the same Finch and Skea metric and the
solution satisfies all the energy conditions including
 TOV-equations. We also check the mass-radius relation, stability  and surface red-shifts for our model and found that their
 behavior is well behaved.\\

The plan of the present paper is as follows: In Sec II we have provided the basic equations in connection to the proposed model
 for strange star of Finch and Skea metric.In Sec. III we dealt with the physical behavior of the star.
Anisotropic behavior, Matching conditions, TOV equations, Energy
conditions, Stability  and Mass-Energy relations are discussed in
different Sub-sections. Concluding remarks are made in Sec. IV.

\section{Quintessence star model}
We assume that the interior space-time of a quintessence star is described by the metric
\begin{equation}
ds^2 = -e^{\nu(r)}dt^2 +  \left(1+\frac{r^2}{R^2}\right)dr^2 +r^2
(d\theta^2 +sin^2\theta d\phi^2), \label{eq1}
\end{equation}
where the metric function $\nu(r)$ is yet to be determined and $R$ is a constant. Consideration of such a space-time, describing
 a paraboloidal geometry, is not new in the analysis of relativistic compact stars\citep{Finch and Skea1989}.

Now we consider a compact star which contains a quintessence field and
a normal matter producing an anisotropic pressure.  Therefore, the Einstein equations can be written as
\begin{equation}
    G_{\mu\nu}=   8 \pi  (   T_{\mu\nu}+  \tau_{\mu\nu}),
         \label{Eq3}
          \end{equation}
where $\tau_{\mu\nu}$ represents the energy momentum tensor of the
quintessence-like field, which is characterized by a free
parameter $\omega_q$ with the restriction
$-1<\omega_q<-\frac{1}{3}$.  According to Kiselev \citep{Kiselev03},
the components of this tensor necessarily satisfy the conditions
of additivity and linearity.  Considering  the
different signatures used in the line elements, the
components can be stated as follows:

\begin{equation}
              \tau_t^t=    \tau_r^r = -\rho_q,
         \label{Eq3}
          \end{equation}
\begin{equation}
              \tau_\theta^\theta=    \tau_\phi^\phi = \frac{1}{2}(
              3\omega_q+1)\rho_q.
         \label{Eq3}
          \end{equation}
However, the most general energy momentum tensor
compatible with spherical  symmetry is
\begin{equation}
               T_\nu^\mu=  ( \rho + p_t)u^{\mu}u_{\nu}
    - p_t g^{\mu}_{\nu}+ (p_r -p_t )\eta^{\mu}\eta_{\nu}
         \label{Eq3}
          \end{equation}
with $u^{\mu}u_{\mu} = -1 $.

Einstein's field equations for the metric (1) accordingly are obtained as ($ c=1,G=1$)
\begin{eqnarray}
~~~~8\pi  (\rho + \rho_q) &=& \frac{1}{R^2}\left(3+\frac{r^2}{R^2}\right)\left(1+\frac{r^2}{R^2}\right)^{-2},\label{eq2}~~\\
~~~~~~8\pi  (p_r - \rho_q) &=& \left(1+\frac{r^2}{R^2}\right)^{-1}\left[\frac{\nu^\prime}{r}+\frac{1}{r^2}\right] - \frac{1}{r^2},
~~\label{eq3}\\
8\pi  \left(p_t +\frac{(3\omega_q+1)}{2}
\rho_q\right)&=& \left(1+\frac{r^2}{R^2}\right)^{-1}\left[\frac{\nu^{\prime\prime}}{2}+\frac{\nu^\prime}{2r}+\frac{{\nu^\prime}^2}
{4}\right]\nonumber \\
&&-\frac{1}{R^2}\left(1+\frac{r^2}{R^2}\right)^{-2}\left[1+\frac{\nu^\prime r}{2}\right]. \label{eq4}
\end{eqnarray}
To solve the above set of equations, we assume that the radial
pressure of the stellar body is proportional to the matter density, i.e.
\begin{equation}
                p_r = m \rho
         \label{Eq4}
          \end{equation}
where m ($>$ 0) is the equation of state parameter.\\
Now, from the metric(1) and equations (6)-(9), we get
\begin{eqnarray}
\nu &=&
\frac{m}{2}\left[2ln\left(1+\frac{r^2}{R^2}\right)+\frac{r^2}{R^2}\right]+
ln\left(\frac{1}{R}\right)+\frac{r^2}{2R^2} \nonumber
\\&&- 8\pi
(1+m)\int r\left(1+\frac{r^2}{R^2}\right)\rho_q dr,\label{eq5}
\end{eqnarray}
Now, we take the quintessence field as
\begin{equation}
                \rho_q = A (2b-r) e^r
         \label{Eq6}
          \end{equation}
where A is a constant of dimension $Length^{-3}$  and b is the
radius of the star. \\

 Although the quintessence field has a repulsive nature, the real
matter at the interior region of the stellar body dominates.  When
we go beyond from the centre of the star then quintessence field
gradually increases. However, up to  the boundary   real matter
dominates and stabilize the stellar structure (Fig.~1).

 By taking the above consideration of the quintessence field we
get
\begin{eqnarray}
\nu &=&
\frac{m}{2}\left[2ln\left(1+\frac{r^2}{R^2}\right)+\frac{r^2}{R^2}\right]+
ln\left(\frac{1}{R}\right)+\frac{r^2}{2R^2} \nonumber
\\&& +8\pi
(1+m)A e^r[(r^2-2br-2r+2b+2)\nonumber \\&& +\frac{1}{R^2}
\{r^4-2(b+2)r^3 \nonumber \\&&+6(b+2)r^2-12(b+2)r+12(b+2) \} ]
\end{eqnarray}

 \begin{equation}
 8\pi \rho = \frac{1}{ R^2}
\frac{\left(3+\frac{r^2}{R^2}\right)}{\left(1+\frac{r^2}{R^2}\right)^2}
-8\pi  A(2b-r)e^r ,\label{eq9}
\end{equation}
\begin{equation}
8\pi  p_{r} = \frac{1}{ R^2}
\frac{m\left(3+\frac{r^2}{R^2}\right)}{\left(1+\frac{r^2}{R^2}\right)^2}
-8\pi  mA(2b-r)e^r ,\label{eq10}
\end{equation}

\begin{eqnarray}
8\pi  p_t &=& \left(1+\frac{r^2}{R^2}\right)^{-1}\left[\frac{\nu^{\prime\prime}}{2}+\frac{\nu^\prime}{2r}+
\frac{{\nu^\prime}^2}{4}\right]\nonumber \\
&&-\frac{1}{R^2}\left(1+\frac{r^2}{R^2}\right)^{-2}\left[1+\frac{\nu^\prime
r}{2}\right]\nonumber \\ &&-8\pi A(2b-r)e^r\frac{(3\omega_q+1)}{2}
\end{eqnarray}
%\begin{eqnarray}
%\frac{dp_r}{dr} &=& - \frac{r(5+\frac{r^2}{R^2})}{12\pi R^4( 1+\frac{r^2}{R^2})^3} < 0,\nonumber\\
%\frac{dp_r}{dr}(r=0) &=& 0,\nonumber\\
%\frac{d^2 p_r}{dr^2} &=& -\frac{5}{12\pi R^4} < 0,\nonumber
%\end{eqnarray}

%showing that the radial pressure also decreases from the centre towards the boundary.
% Thus, the energy density and the radial pressure are well behaved in the interior of the stellar configuration.
% Variations of the energy-density and two pressures have been shown in Fig.~(\ref{fig1}) and (\ref{fig2}), respectively.

%The anisotropic parameter $\Delta (r)  = \frac{2}{r}\left(p_t-p_r\right)$ representing the anisotropic tress is obtained as

\begin{figure}[htbp]
    \centering
        \includegraphics[scale=.3]{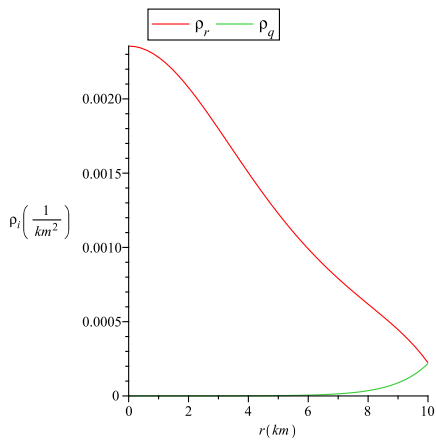}
        \caption{ Density comparison of real and quintessence matters  at the
stellar interior.}
    \label{fig:1}
\end{figure}

\section{Physical Analysis}
We note that the physical behavior of the star depends on the constants R and $\rho_q $. Here, we need
to put appropriate assumption to get the realistic model.In this section we will discuss the following features of our model :
\subsection{Anisotropic Behavior of the star}

From Eq.~(\ref{eq9}), the central and surface densities are
obtained as
\begin{eqnarray}
\rho_0 &=& \frac{3}{8\pi  R^2}- 2Ab  ,\nonumber\\
\rho_b &=& \frac{1}{8\pi
R^2}\left(3+\frac{b^2}{R^2}\right)\left(1+\frac{b^2}{R^2}\right)^{-2}-A b e^b\nonumber
\end{eqnarray}
where we have assumed that $b$ is the radius of the star and density
due to quintessence field at the centre of the star, $\rho_q$ = 2Ab.
Therefore, We can say that
\begin{equation}
\rho_{0eff} = \frac{3}{8\pi  R^2} > 0
\end{equation}
and
\begin{equation}
 \rho_{eff} = \frac{1}{8\pi  R^2}
\frac{\left(3+\frac{r^2}{R^2}\right)}{\left(1+\frac{r^2}{R^2}\right)^2}
> 0
\end{equation}
Now, we check whether at the centre the effective matter density
dominates or not. Here, we see that
\begin{eqnarray}
\frac{d\rho_{eff}}{dr} &=& - \frac{r(5+\frac{r^2}{R^2})}{4\pi  R^4( 1+\frac{r^2}{R^2})^3}< 0,\nonumber\\
\frac{d\rho_{eff}}{dr} (r=0) &=& 0,\nonumber \\
\frac{d^2 \rho_{eff}}{dr^2}(r=0) &=& -\frac{5}{4\pi  R^4} <
0.\nonumber
\end{eqnarray}
Clearly, at the centre, the effective density of the star is maximum
and it decreases radially outward. Similarly, from Eq.(7), we get
\begin{eqnarray}
\frac{dp_{reff}}{dr} &=& - \frac{mr(5+\frac{r^2}{R^2})}{4\pi  R^4( 1+\frac{r^2}{R^2})^3} < 0
\end{eqnarray}
Here, again at the centre (r=0),
\begin{eqnarray}
\frac{dp_{reff}}{dr}(r=0) &=& 0\nonumber\\
\frac{d^2 p_{reff}}{dr^2}(r=0) &=& -\frac{5m}{4\pi  R^4} < 0\nonumber
\end{eqnarray}
Therefore, at the centre, we also see that the effective radial pressure is maximum  and it decreases from the centre
towards the boundary. Thus, the effective energy density and the effective radial
pressure are well behaved in the interior of the stellar
structure. Variations of the effective energy-density and effective radial pressure
have been shown in Fig.~2 and Fig.~3, respectively.

The anisotropic parameter $\Delta (r)  =
\frac{2}{r}\left(p_t-p_r\right)$ representing the anisotropic stress
is shown in Fig.~4.

\begin{figure}[htbp]
    \centering
        \includegraphics[scale=.3]{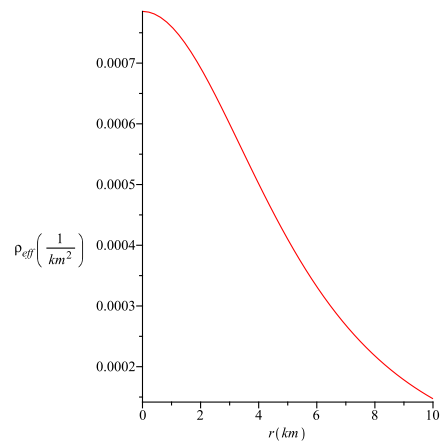}
       \caption{ Variation of the effective energy-density($\rho$) at the
stellar interior.}
    \label{fig:2}
\end{figure}
\begin{figure}[htbp]
    \centering
        \includegraphics[scale=.3]{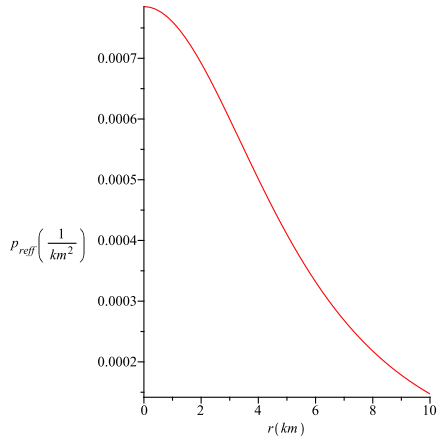}
        \caption{ Variation of the effective radial pressure ($p_r$) at the
stellar interior.}
    \label{fig:3}
\end{figure}
\begin{figure}[htbp]
    \centering
        \includegraphics[scale=.3]{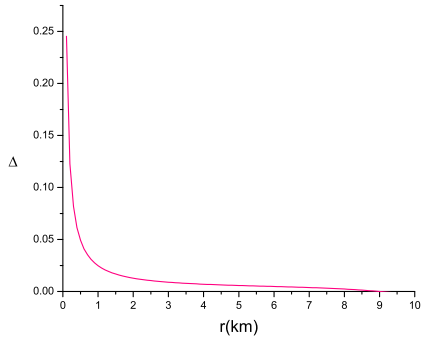}
        \caption{ Effective anisotropic behavior $\Delta (r)$  at the
stellar interior.}
    \label{fig:4}
\end{figure}

\subsection{Matching Conditions}
Interior metric of the star should be matched to the Schwarzschild exterior metric at the boundary ($r=b$).
\begin{equation}
ds^2 = - \left(1-\frac{2M}{r}\right)dt^2 +  \left(1-\frac{2M}{r}\right)^{-1}dr^2 +r^2
(d\theta^2 +sin^2\theta d\phi^2), \label{eq1}
\end{equation}
Assuming the continuity of the metric functions $g_{tt}$,$g_{rr} $and$\frac{\partial g_{tt}}{\partial r}$ at the boundary, we get

\begin{eqnarray}
\left(1+\frac{b^2}{R^2}\right)^{-1} &=& 1 - \frac{2M}{b},\label{eq13}\\
\nu(r=b) &=& \ln\left(1-\frac{2M}{b}\right) = \frac{m}{2}\left[2\ln(1+\frac{b^2}{R^2})+\frac{b^2}{R^2}\right]\nonumber \\
&&+\ln(\frac{1}{R})+\frac{b^2}{2R^2}-8\pi (1+m)Ae^b\nonumber \\
&&\left[ (b^2-2)+\frac{1}{R^2}\left(b^4-2b^3+12b-24\right) \right] .\label{eq14}
\end{eqnarray}
From Eq.~(\ref{eq13}) , we get the compactification factor as
\begin{equation}
\frac{M_{eff}}{b} = \frac{b^2}{2R^2}\left(1+\frac{b^2}{R^2}\right)^{-1}.\label{eq15}
\end{equation}
Equation ~(\ref{eq14}) yields the value of unknown constant A in
terms of compactification factor, equation of state parameter  and
b, the radius of the star.

\subsection{TOV equation}
For an anisotropic fluid distribution, the generalized TOV equation has the form
\begin{equation}
\frac{dp_{r~eff}}{dr} +\frac{1}{2} \nu^\prime\left(\rho_{eff}
 + p_{r~eff}\right) + \frac{2}{r}\left(p_{r~eff} - p_{t~eff}\right)
= 0.\label{eq18}
\end{equation}
Following \citet{Leon1993}, we write the above equation as
\begin{equation}
-\frac{M_G\left(\rho_{eff}+p_{r~eff}\right)}{r^2}e^{\frac{\lambda-\nu}{2}}-\frac{dp_{r~eff}
}{dr}
 +\frac{2}{r}\left(p_{t~eff}-p_{r~eff}\right) = 0, \label{eq19}
\end{equation}
where $M_G(r)$ is the effective gravitational mass inside a sphere of radius $r$ and is given by
\begin{equation}
M_G(r) = \frac{1}{2}r^2e^{\frac{\nu-\lambda}{2}}\nu^{\prime}.\label{eq20}
\end{equation}
where $e^{\lambda(r)} = 1+\frac{r^2}{R^2} $

which can be derived from the Tolman-Whittaker formula and the Einstein's field equations. The modified TOV equation describes
 the equilibrium condition for the compact quintessence star subject to
effective gravitational($F_g$) and effective hydrostatic($F_h$) plus another force due to the effective anisotropic($F_a$)
 nature of the stellar object as
\begin{equation}
F_g+ F_h + F_a = 0,\label{eq21}
\end{equation}
where,
\begin{eqnarray}
F_g &=& -\frac{1}{2}\left(\rho_{eff}+p_{r~eff}\right)[\frac{m}{R^2}\frac{3r + \frac{r^3}{R^2}}{1+ \frac{r^2}{R^2}}+ \frac{r}{R^2} \nonumber \\
&& - 8 \pi A(1+m)e^r\left(2br-r^2+\frac{2br^3}{R^2}-\frac{r^4}{R^2}\right)]\label{eq22}\\
F_h &=& -\frac{dp_{r~eff}}{dr} \label{eq23}\\
F_a &=& \frac{2}{r}\left(p_{t~eff} -p_{r~eff}\right)\label{eq24}
\end{eqnarray}
We plot ( Fig. 5 )   the behaviors of pressure anisotropy,
gravitational and hydrostatic forces at the stellar interior of 4U
1820-30 which indicates  the static equilibrium configurations do
exist in the presence of pressure anisotropy, gravitational and
hydrostatic forces.
%\begin{figure}
 %     \plotone{F.eps} \caption{Three different forces acting on fluid
%elements in static equilibrium is shown against $r$. \label{fig4}}
%\end{figure}

\begin{figure}[htbp]
    \centering
        \includegraphics[scale=.3]{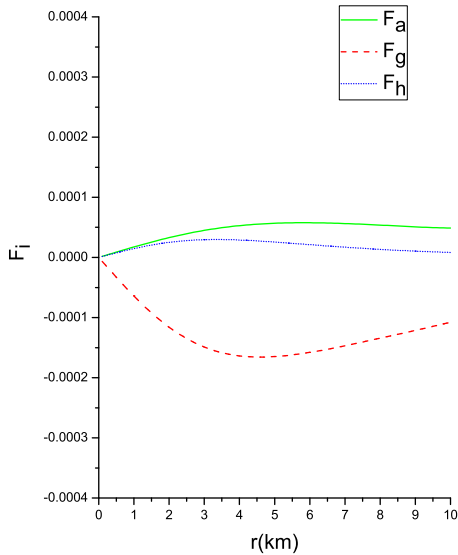}
        \caption{ Behaviors of pressure anisotropy, gravitational and hydrostatic forces at the
stellar interior of 4U 1820-30.}
    \label{fig:5}
\end{figure}

\subsection{Energy conditions}
All the energy conditions, namely, null energy condition(NEC), weak energy condition(WEC), strong energy condition(SEC)
and dominant energy condition(DEC), are satisfied. In our model, we get the following energy conditions at the centre ($r=0$) :\\
(i) NEC: $p_{0~eff}+\rho_{0~eff}\geq0$ ,\\
(ii) WEC: $p_{0~eff}+\rho_{0~eff}\geq0$  , $~~\rho_{0~eff}\geq0$  ,\\
(iii) SEC: $p_{0~eff}+\rho_{0~eff}\geq0$  ,$~~~~3p_{0~eff}+\rho_{0~eff}\geq0$ ,\\
(iv) DEC: $\rho_{0~eff} > |p_{0~eff}| $.
\subsection{Stability}
For a physically acceptable model, one expects that the velocity of
sound should be within the range  $0 \leq  v_s^2=(\frac{dp}{d\rho})
\leq 1$\citep{Herrera1992,Abreu2007}.   In our case, we have
\begin{equation}
 v_{sr}^{2} = \frac{1}{3}.\label{eq25}
\end{equation}
We plot the radial and transverse sound speeds in Fig.6 which shows
that these parameters satisfy the inequalities $0\leq v_{sr}^2 \leq
1$ and $0\leq v_{st}^2 \leq 1$ everywhere within the stellar object.
We also note that $v^2_{st}-v^2_{sr}\leq 1$. Since, $0\leq v_{sr}^2
\leq 1$ and $0\leq v_{st}^2 \leq 1$, therefore,  $\mid v_{st}^2 -
v_{sr}^2 \mid \leq 1 $. In Fig.7, we have plotted  $\mid v_{st}^2 -
v_{sr}^2 \mid$. According to Herrera's \citet{Herrera1992} cracking
(or overturning) theorem, these results show that the star model is
stable.  [  Herrera's \citet{Herrera1992} cracking (or overturning)
theorem  states that the region  for which radial speed of sound is
greater than the transverse speed of sound is a potentially stable
region. ]

\begin{figure}[htbp]
   \centering
        \includegraphics[scale=.3]{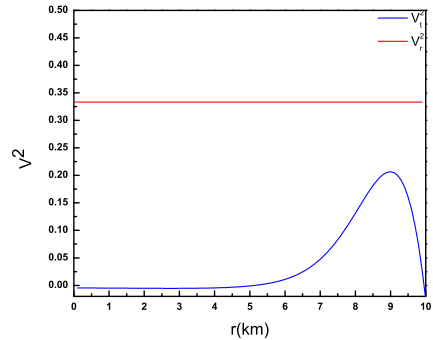}
       \caption{ Variation of the radial and transverse sound speed of 4U 1820-30.}
    \label{fig:6}
\end{figure}

\begin{figure}[htbp]
    \centering
        \includegraphics[scale=.3]{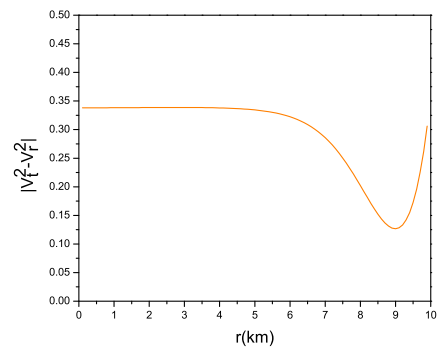}
       \caption{ Variation of $v_{st}^2 - v_{sr}^2$ of strange star 4U 1820-30.}
    \label{fig:7}
\end{figure}

\subsection{Mass-Radius relation}
In this section, we study the maximum allowable mass-radius ratio
in our model. According to Buchdahl \citep{Buchdahl1959}, for a static spherically symmetric perfect fluid
allowable mass-radius ratio is given by $\frac{2M}{R} <
\frac{8}{9}$ for a more generalized expression for the same see
mass in terms of the  effective energy density $\rho_{eff}$ can be expressed as
\begin{equation}
\label{eq34}
 M_{eff}=4\pi\int^{b}_{0} \rho_{eff}~~ r^2 dr =
 \frac{b}{2}\left[\frac{\frac{b^2}{R^2}}{1+\frac{b^2}{R^2}}\right]
\end{equation}

We note that a
constraint on the maximum allowed mass-radius ratio in our case is
similar to the isotropic fluid sphere, i.e., $\frac{mass}{radius}
< \frac{4}{9}$ as obtained earlier. The compactness of the star is
given by
\begin{equation}
\label{eq35} u= \frac{ M_{eff}(b)} {b}=
\frac{1}{2}\left[\frac{\frac{b^2}{R^2}}{1+\frac{b^2}{R^2}}\right]
\end{equation}
\begin{figure}[htbp]
    \centering
        \includegraphics[scale=.3]{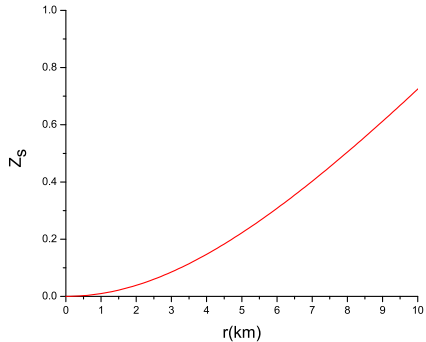}
        \caption{ Variation of the red-shift function of strange star 4U 1820-30}
    \label{fig:8}
\end{figure}

The surface red-shift ($Z_s$) corresponding to the above
compactness ($u$) is obtained as
\begin{equation}
\label{eq36} 1+Z_s= \left[ 1-(2 u )\right]^{-\frac{1}{2}} ,
\end{equation}
where
\begin{equation}
\label{eq37} Z_s= \sqrt{1+\frac{b^2}{R^2}}-1
\end{equation}
Thus, the maximum surface red-shift for the quintessence anisotropic stars of
different radius could be found very easily ( Fig. 8 ).
\section{Conclusion}
In the present work, we investigate the nature of the quintessence
type star by taking the Finch-Skea metric which describes the
paraboidal geometry. In general compact stars, due to their high
density, becomes anisotropic in nature. That's why we consider the
anisotropic behavior of the star to make more generalized model.As
the structure of these star are till not known, we assume that the
interior stellar structure may constituted  not only ordinary matter
but also with a quintessence dark energy having a characteristic
parameter
 $\omega_q$ such that $-1 < \omega_q < -\frac{1}{3}$. The quintessence dark energy seems to be  responsible
 for the accelerated expansion of the universe, therefore, inclusion of the quintessence dark
 energy in the interior of the highly compact stars is justified.
  The presence of dark energy
  motivates us to consider the existing strange stars are a mixture of both ordinary
  matter and quintessence dark energy in different proportions. We all know that quintessence
   has a repulsive nature. But, the real matter at the central region of the stellar body dominates
    to make the effective energy density to be positive. But when we go beyond from the centre of the star then quintessence field
gradually increases. However, up to  the boundary   real matter
dominates and stabilize the stellar structure. In conclusion, we can say that
    incorporation of quintessence matter with the real one describes the well-known compact stars (e.g.
    neutron stars,white dwarf stars, 4U 1820-30, Her X-1, SAX J 1808.4-3658 etc.) in a good manner in all
     respects.\\

\section*{Acknowledgments} MK, FR gratefully acknowledge support
 from IUCAA, Pune, India under Visiting Associateship under which a part
  of this work was carried out. SMH is  thankful to IUCAA also for giving him an opportunity to visit IUCAA where a part
   of this work was carried out. FR is also thankful to UGC for providing financial support under Research Award Scheme.
We are thankful to the  referee for  informing some of the important
references.

\end{document}